\documentclass[reprint,
superscriptaddress,
 amsmath,amssymb,amsfonts,
 aps,
 prl,
]{revtex4-1}
\usepackage[hidelinks]{hyperref}
\usepackage{graphicx}
\usepackage{dcolumn}
\usepackage{bm}
\usepackage{comment}

\begin{document}

\title{A Very Effective and Simple Diffusion Reconstruction for the Diluted Ising Model}

\author{Stefano Bae}
 \email{stefano.bae@uniroma1.it}
 \thanks{First and corresponding author}
 \affiliation{Dipartimento di Fisica, Sapienza Universit\`a di Roma, P.~A.~Moro 5, 00185 Roma (Italy)}
\author{Enzo Marinari}
 \email{enzo.marinari@uniroma1.it}
 \thanks{The two last authors contributed equally.}
\author{Federico Ricci-Tersenghi}
 \email{federico.ricci@uniroma1.it}
 \thanks{The two last authors contributed equally.}
 \affiliation{Dipartimento di Fisica, Sapienza Universit\`a di Roma, P.~A.~Moro 5, 00185 Roma (Italy)}
 \affiliation{Nanotec-CNR, Unit\`a di Roma, and INFN-Sezione di Roma1, P.~A.~Moro 5, 00185 Roma (Italy)}

\date{\today}

\begin{abstract}
Diffusion-based generative models are machine learning models that use diffusion processes to learn the probability distribution of high-dimensional data. In recent years, they have become extremely successful in generating multimedia content.  However, it is still unknown whether such models can be used to generate high-quality datasets of physical models. In this work, we use a Landau-Ginzburg-like diffusion model to infer the distribution of a $2D$ bond-diluted Ising model. Our approach is simple and effective, and we show that the generated samples correctly reproduce the statistical and critical properties of the physical model.
\end{abstract}

\maketitle

\paragraph{Introduction}
Diffusion-based generative models (DM) are machine learning models designed to learn the probability distribution of high-dimensional data and use stochastic differential equations to generate new samples under such distribution \cite{sohl, comprehensiveSurvey,song2020score,song21}. They are widely applied in high-quality image and video generation \cite{song2020score, ddpmHo}, and recently, together with other generative models, DMs are also finding applications in improving simulations of physical models \cite{li24, dehaan21,devlin24} and enhancing Monte Carlo simulations for low-dimensional functions \cite{hunt23}. There have been several efforts to rephrase these models in the statistical mechanics framework and to improve our theoretical understanding of their behavior. Relations with spontaneous symmetry breaking and phase transitions \cite{biroli23, ambrogioni2023, raya2024spontaneous} have been established, the characterization of time regimes \cite{biroli24} and the connection with associative memory \cite{AssociativeAL} have been analyzed. Also, the physical limits of the method and advantages and disadvantages with respect to standard Monte Carlo and Langevin sampling have been analyzed \cite{ghio}.

As the name suggests, DMs use a diffusion process to accomplish such tasks. The force field of the diffusion process that recursively transforms the initial data into a random noise (typically a Gaussian white noise) is a priori not known, and the relevant goal is to learn it. If one can do that, a time-reversed diffusion process can then be used to generate a new sample starting from what appears as pure noise. This task is usually achieved by the use of neural networks. We propose here a new method based on an approximate score function that allows us to reconstruct very efficiently the diluted Ising model (DIM). Disordered models were not studied before with DM, so it is a crucial test if a reconstruction method can analyze the DIM. We will give this question a positive answer, and we will argue that an approach based on a simple structure can be very effective. We believe that these findings have a relevant impact on the more general question of the diffusion-based analysis. 

We start from an initial training dataset $\{\vec{a}^\mu\}_{\mu =1}^{P}$. The $P$ elements of the set are $N$ dimensional vectors, and they are distributed according to an unknown initial distribution $P_0(\vec{a})$. These vectors evolve under an Ornstein-Uhlenbeck diffusion process, and the corresponding stochastic equation in the Ito formulation is:
\begin{equation}
    \frac{dx_i(t)}{dt} = -x_i(t) + \eta_i(t) \;,
    \label{eqn:forward}
\end{equation}
where $\eta_i(t)$ is a white Gaussian noise with correlation $\langle \eta_i(t)\eta_j(t')\rangle = 2\delta_{ij}\delta(t-t') $. The stationary distribution of this process is normal $\mathcal{N}(\vec{x};\vec{0},\mathbb{I})$ centered at the origin and with an identity covariance matrix. The probability distribution of this forward process is
\begin{equation}
    P\left(\vec{x},t\right) = \int P_0\left(\vec{a}\right)\; \mathcal{N}\left(\vec{x};\vec{a}e^{-t},\Delta\left(t\right)\mathbb{I}\right)d\vec{a} \;,
    \label{eqn:forw_prob}
\end{equation}
where the mean of the normal distribution is $\vec{a}e^{-t}$ and the variance is $\Delta(t) = 1-e^{-2t}$. 
Ideally, given this $P(\vec{x},t)$, one can derive the ideal force field, namely the \textit{score function}
\begin{equation}
    \mathcal{F}_i(\vec{x},t) =  \frac{\partial  \log P(\vec{x},t)}{\partial x_i}=  -\frac{x_i - \langle a_i \rangle_{\vec{x}} \,e^{-t}}{\Delta(t)} \;,
    \label{eqn:truescore}
\end{equation}
where $\langle a_i \rangle_{\vec{x}}$  is the average over the conditional probability $P(\vec{a}|\vec{x},t)$
and $\mathcal{F}_i(\vec{x},t)$ can be used to define the backward process that will generate new samples \cite{anderson82,song21}:
\begin{equation}
    -\frac{dy_i}{dt} = y_i +2T\mathcal{F}_i(\vec{y},t) +\eta_i(t) \;.
    \label{eqn:backward}
\end{equation}
However, the distribution $P_0(\vec{a})$ of real-world data is not known, so in realistic applications, the score function must be learned from the data using a suitable approach (typically a deep neural network).

In our work, we choose to approximate the forward process distribution $P(\vec{x},t) \approx Q(\vec{x},t;\theta)$ with a function to be discussed in detail below, where $\theta$ are generic parameters. The approximated score is given by
\begin{equation}
    S_i(\vec{x},t; \theta) \equiv \frac{\partial  \log Q(\vec{x},t;\theta)}{\partial x_i}\;.
    \label{eqn:approxscore}
\end{equation}
Since we want to approximate the true score (\ref{eqn:truescore}) with (\ref{eqn:approxscore}), we minimize the mean squared loss to find the optimal value of the parameters $\theta$:
\begin{equation}
    \mathcal{L}\left(\theta\right) =
    \biggl< \sum_{k=1}^N \Big(S_k(\vec{x},t;\theta) - \mathcal{F}_k(\vec{x},t) \Big)^2
    \biggr>_{\vec{x},\vec{a}} \;,
    \label{eqn:mseloss}
\end{equation}
where $\langle \cdot \rangle_{\vec{x},\vec{a}} =\int (\cdot )P_0\left(\vec{a}\right)\mathcal{N}\left(\vec{x};\vec{a}e^{-t},\Delta\left(t\right)\mathbb{I}\right) \,d\vec{x}d\vec{a}$.

Here we try, on the DIM, a very simple approach, that does not need any deep learning system, and we show that it works and that it is very effective. We define $\log Q(\vec{x},t)$ as a quadratic term plus a fourth-order term, and we obtain in this way a non-linear score function that is able to reproduce the statistical and critical properties of the DIM. This choice of the forward probability distribution, which is strongly reminiscent of a Landau-Ginzburg free energy, is able to generate the correct clustering of the data, reproducing the correct magnetization statistics. Analyzing the spatial correlation function and the magnetic susceptibility we find further and strong evidence that the generated samples show the same critical behavior as the original model. Finally, in light of \cite{biroli24}, we discuss the generalization and the memorization regimes, and we characterize the speciation time, i.e.\ the time at which the trajectories start to fall in one of the free energy minima.

\paragraph{The diluted Ising model}
We consider a link diluted Ising model on a $2D$ square lattice of linear size $L$ and volume $N = L^2$. Given $N$ Ising variables, $a_i = \pm 1$, the DIM Hamiltonian is defined as
\begin{equation}
    \mathcal{H}(\vec{a}) = -\sum_{i,j}J_{ij} a_i a_j\;,
    \label{eqn:dim_ham}
\end{equation}
where the couplings $J_{ij}$ are null if $|i-j| \neq 1$ and distributed according to $p\,\delta(J_{ij}-1) + (1-p)\,\delta(J_{ij})$ if $|i-j| = 1$. The bond concentration $p$ controls the probability of having a bond between nearest neighbors, and in the limit of $p \rightarrow 1 $ one recovers the usual $2D$ Ising Model. A set of i.i.d.\ $\left\{J_{ij}\right\}$ extracted according to this probability distribution defines a quenched realization of the disorder, i.e.\ a disorder sample.

At equilibrium, the configurations of the variables $\vec{a}$ are distributed according to the Boltzmann-Gibbs measure
\begin{equation}
    \label{eqn:pgb}
    P_{BG}(\vec{a}) = \frac{1}{\mathcal{Z}} e^{-\beta \mathcal{H}(\vec{a})}\;,
\end{equation}
where the partition function $\mathcal{Z}$ normalizes $P_{BG}(\vec{a})$, and $\beta = 1/T$ is the inverse temperature of the system. The order parameter that characterizes the different thermodynamic phases is the magnetization
\begin{equation}
    m = \frac{1}{N}\sum_{i=1}^{N} a_i \;.
    \label{eqn:magn}
\end{equation}
The value of the critical temperature $T_c(p)$ that separates the two phases depends on the bond concentration.

Our goal is to approximate (\ref{eqn:pgb}) and generate samples using a DM, to reproduce correctly the main properties of the DIM, such as the critical exponents and the distribution of the magnetization $P(m)$. The DIM has been analyzed in detail, and there exists a large amount of results characterizing its critical and thermodynamic behavior \cite{griff69,harris74,ballesteros97}; this is very useful to fully characterize the consistency of our results. 

\paragraph{Our method}
The typical (and powerful) attitude adopted to reconstruct an effective score function that will govern the DM is to use a strongly over-parametrized deep network, for example, U-Nets \cite{ddpmHo}. This can be a very useful approach, but the fine-tuning and training of such a deep network can be cumbersome. It is natural to wonder if this is really needed. Even if reconstructing a bona fide statistical model with quenched disorder is not at all a trivial task, we focus on the possibility, that we judge valuable, to use a simple effective score function, and to succeed in the reconstruction in this way. We do indeed succeed, and in this way collect positive results, since we are able to backward reconstruct a non-trivial model with quenched disorder by using a very simple parametrization. 

The simplest form one could select for implementing the forward process distribution would be Gaussian. This cannot lead to success: the score obtained in this way is linear, and once it is plugged in the backward Langevin process (\ref{eqn:backward}) it can only allow to obtain Gaussian distributed samples \cite{biroli23}. To try to overcome this problem, we use the simplest possible approach, and we add a fourth-order term to the Gaussian distribution, by setting
\begin{equation}
    \log Q(\vec{x},t) =  -\frac{1}{2} \sum_{i,j}^{1,N} x_i A_{ij} x_j - \frac{g}{4N^2}\left(\sum_{i=1}^{N} x_i\right)^4 + \mathcal{C}\;,
    \label{eqn:log_forw_prob}
\end{equation}
where $\bm{A}$ is a symmetric square matrix of dimension $N$ and $g$ is the parameter that controls the effects of the fourth-order term. $\mathcal{C}$ is a normalization constant. The added term is the global magnetization, scaled by a factor $\sqrt{N}$ and raised to the fourth power. The score corresponding to this forward distribution is:
\begin{equation}
    \frac{\partial  \log Q(\vec{x},t)}{\partial x_k} = -
    \sum_{j = 1}^N A_{kj}x_j - \frac{g}{N^2} \left(\sum_{i=1}^{N} x_i\right)^3\;.
    \label{eqn:ourscore}
\end{equation}
We can intuitively see that the term $\left(\sum_{i=1}^{N} x_i\right)^3$ will help the variables to correlate in a coherent way. In fact, each sum is just the magnetization of the system and if we display it as $\left(\sum_{i=1}^{N} x_i\right)\left(\sum_{i=1}^{N} x_i\right)^2$ we can see that the squared sum gives a measure of how data is correlated, while the first term also accounts for the direction of the $x_i$. In other words, if the variables exhibit correlations during the backward process, this term will amplify them by acting as an external field. Moreover, the added quartic term is the smallest power that can break the symmetry. Indeed, the addition of higher order powers does not significantly improve the performance of the model.

After the process of minimization of (\ref{eqn:mseloss}), which is described in detail in the Supplemental Information (SI), we obtain the following expressions for the matrix $\bm{A}$ and the parameter $g$:
\begin{equation}
    (\bm{A}(t) )_{lm} = (\bm{C}(t)^{-1})_{lm} - \frac{g(t)}{N^2} \sum_{j=1}^{N} \omega^{(4)}_j(t) (\bm{C}(t)^{-1})_{jm}\;,
    \label{eqn:eqnA}
\end{equation}
\begin{equation}
    g(t) = N \frac{ \frac{\omega^{(4)}(t)- \Tilde{\omega}^{(4)}(t)}{\Delta(t)} - \gamma^{(0,4)}(t)}{\omega^{(6)}(t)-\gamma^{(4,4)}(t)}\;,
    \label{eqn:eqng}
\end{equation}
where the matrix $\bm{C}(t)$ is the correlation matrix of the forward process, which depends on the correlation of the training data $\bm{C}_0$:
\begin{equation}
    (\bm{C}(t))_{km}=\langle x_k x_m \rangle_{\vec{x},\vec{a}} = (\bm{C}_0)_{km}e^{-2t} + \Delta(t) \delta_{km}
    \label{eqn:corr}
\end{equation}
We notice that the first term in (\ref{eqn:eqnA}), at time $t=0$, is the mean-field inference of the couplings $\beta J_{ij}$ \cite{zecchina17}.
The other terms, $\omega^{(4)}(t)$, $\omega^{(6)}(t)$, $\Tilde{\omega}^{(4)}(t)$, $\omega_j^{(4)}(t)$, $\gamma^{(0,4)}(t)$ and $\gamma^{(4,4)}(t)$, are moments and combinations of moments of the forward process, whose details can also be found in the SI. They depend only on $\bm{C}_0$ and on the single sample initial magnetization $m_0(\vec{a}^\mu)$.

\begin{figure}
    \centering
    \includegraphics{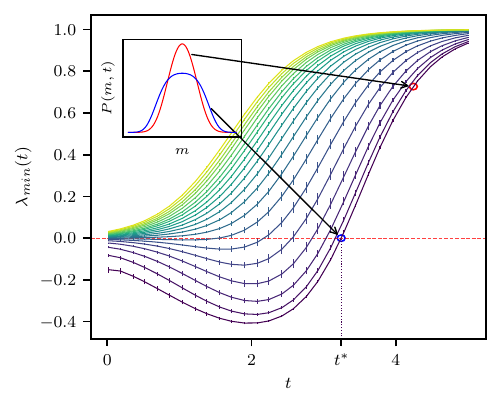}
    \caption{The smallest eigenvalue $\lambda_{min}(t)$ of $\bm{A}(t)$ for the DIM with $L=50$ and $p=0.8$. Each line is averaged over $20$ samples. The different colors for different lines correspond to different temperatures linearly spaced in the range $T\in [1.600,2.075]$. Darker lines are for lower values of $T$. For higher values of $T$, $\lambda_{min}(t)$  never touches the $0$ line, in agreement with the fact that for $T>T_c$ there is only a paramagnetic minimum. When $T$ decreases $\lambda_{min}(t)$  crosses the x-axis, because of the presence of two minima. The two inset plots correspond to the distribution of the magnetization in time $P(m,t)$ for $t>t^*$ (red one) and $t=t^*$ (blue one).}
    \label{fig:minlambda}
\end{figure}

\paragraph{Results}
We have generated the training data for the DIM using the Wolff algorithm, an improved cluster Monte Carlo Markov Chain \cite{wolff89,newmanb99}. We have used dilution $p=1.0$ for studying the pure Ising model and $p=0.8$ for analyzing a diluted model. In all cases, the DM was trained with $P_{train} = 20000$ independent samples \footnote{The Wolff algorithm has a small correlation time \cite{newmanb99} and to generate independent samples, we took measurements that were appropriately spaced in time.}. We eventually generated $P_{gen} = 10000$ samples. Once training samples have been generated, the parameters $\bm{A}(t)$ and $g(t)$ can be estimated using (\ref{eqn:eqnA}) and (\ref{eqn:eqng}). Then, using (\ref{eqn:backward}) new configurations of the system can be obtained.
We stress that because of the nature of the diffusion process, the generated data are real instead of the original binary ones, and we worked starting from them. We eventually obtained binary spin values by applying a sign function (and a discussion of this procedure is given in the SI).

\begin{figure}
    \centering
    \includegraphics{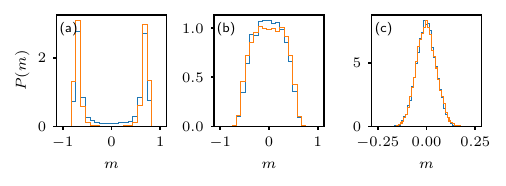}
    \caption{The distribution of the magnetization $P(m)$, with $L=100$ and $p=0.8$. The orange and blue lines correspond respectively to the original and the generated samples. From left to right we display the results for the low $T$ broken phase (a), for the critical point (b), and for the paramagnetic phase (c). The samples generated by backward diffusion match the original distribution.}
    \label{fig:pdf_magn}
\end{figure}

To develop a multimodal $Q(\vec{x},t)$ that correctly reproduces the true $P(\vec{x},t)$, we expect that the Hessian of the effective free energy in (\ref{eqn:log_forw_prob}), $\bm{A}(t)$, should at some point develop a soft mode. Thus, we analyze the smallest eigenvalue of $\bm{A}(t)$, $\lambda_{min}(t)$, for training sets at different temperatures. We show in Fig.~\ref{fig:minlambda} that there exists a time $t^*$ at which the eigenvalue changes sign from positive to negative going backward in $t$. This means that the effective free energy is developing new minima that should correspond to the two minima of the DIM \cite{raya2024spontaneous}. When the diffusion dynamics is at $t>t^*$, all the eigenvalues are positive, indicating that the free energy is convex and $Q(\vec{x},t)$ is unimodal. At $t = t^*$, the free energy becomes flat along the direction corresponding to the eigenvector of $\lambda_{min}(t^*)$, and fluctuations along this direction are enhanced. For $t<t^*$, the uni-modal $Q(\vec{x},t)$ is no longer stable, and spontaneously evolves towards a $Q(\vec{x},t)$ with two peaks. Indeed, looking at the distribution of the magnetization $P(m)$ shown in Fig.~\ref{fig:pdf_magn}, we can see that in different thermodynamic phases --- above, around, and below $T_c(p)$ --- the distribution of the magnetization of the generated samples is in agreement with that of the original data. Particularly remarkable is the observation in panel (a) that the peaks are reproduced correctly in the broken phase, notwithstanding the simplicity of our score function. It is also correct to note that there is a slight abundance of $m\approx0$ configurations.

\begin{figure*}
    \centering
    \includegraphics[width=\textwidth]{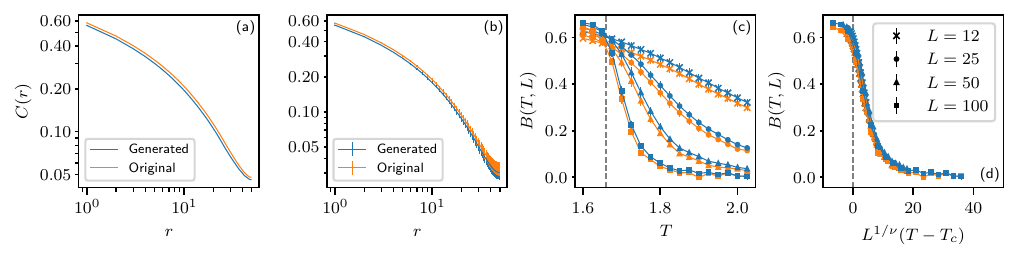}
    \caption{Panels (a) and (b): spatial correlation function $C(r)$ at criticality for the original (orange above) and the reconstructed (blue below) data, with $L=100$. We show the pure Ising model, $p=1.0$, in (a), and the DIM with $p=0.8$ in (b). $C(r)$ has been averaged over $10$ disorder samples. In both cases, the generated correlations show the same critical decay as the original one. Panels (c) and (d): DIM Binder parameter $B(T,L)$ as a function of the temperature $T$, with $L \in\{12, 25, 50, 100\}$ and $p =0.8$, averaged over $10$ disorder samples. The orange points and lines are for the original data, the blue ones are for the generated samples. The vertical dashed line is for the critical temperature $T_c(p=0.8) \simeq 1.650$ \cite{hadjia11}. In panel (c) we show the intersection of the generated and original data at $T_c(p=0.8)$. In panel (d) we show that the values of the Binder parameter rescaled using the critical exponents of the $2D$ Ising universality class collapse onto a single curve.}
    \label{fig:corr_binder}
\end{figure*}

Our main goal is to show that our (simple) diffusion model is able to reconstruct in detail the critical properties of the DIM (and we have already shown in Fig.~\ref{fig:pdf_magn} that it reconstructs accurately the bimodal distribution of $m$ below $T_c$). We analyze next the $x-x$ two points spatial correlation function $C(r)$ and the fourth moment of the magnetization, namely the Binder parameter $B(L,T)$.

The spatial correlation function $C(r)$ is defined as
\begin{equation}
    C(r) \equiv \frac{1}{S_r}\sum_{|i-j|=r} \Big[ \langle x_ix_j \rangle_T - \langle x_i \rangle_T \langle x_j\rangle_T \Big]\;,
\end{equation}
where the sum runs over all pairs of spins at distance $r$ along the axis, being $S_r$ their number. $\langle\cdot \rangle_T$ is the average over the Gibbs measure (\ref{eqn:pgb}) at temperature $T$, while $[\cdot]$ is the average over the disorder (we have used few tens of samples).
The Binder $B(L,T)$ is defined as
\begin{equation}
    B(L,T) \equiv 1-\frac{\left[\langle m^4\rangle_T\right]}{3\left[\langle m^2\rangle_T\right]^2}\;.
\end{equation}

We show in Fig.~\ref{fig:corr_binder} that the correlations that characterize the generated samples at the critical temperature decay as in the original, Boltzmann distributed samples: as one would expect the system is characterized by the same critical exponent $\eta = \frac14$ (but for size-dependent logarithmic corrections \cite{ballesteros97})) of the $2D$ pure Ising model universality class. By analyzing $B(L,T)$ we also checked that the value $\nu = 1$  is reproduced correctly by our backward reconstruction. We show in panels (c) and (d) of Fig.~\ref{fig:corr_binder} the scaling behavior of $B(L,T)$, derived by using the $2D$ Ising exponents (again ignoring logarithmic corrections), that confirms that the critical properties of the generated data reproduce with high accuracy the ones of the original data \cite{ballesteros97}. The intersection of the Binder parameter is used in literature to determine the critical temperature and the critical exponents of the model \cite{binder81}, and we can see that the analysis of the generated data leads to the same estimates as for the original data. 

\begin{figure}
    \centering
    \includegraphics{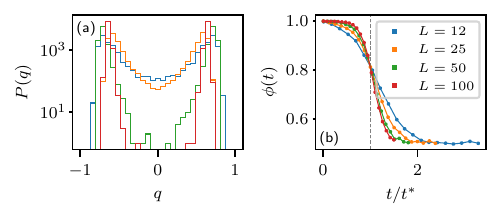}
    \caption{We show in panel (a) the distribution of the overlap between the generated and the original samples in the broken phase $T=1.6<T_c$ with $p=0.8$. The two symmetric peaks correspond to samples that have the same magnetization, while there is no peak at $q=1$ suggesting that the model is typically not generating samples that are in the training set. In panel (b) we show the probabilities that two cloned trajectories at a time $t$ fall in the same peak as a function of $t/t^*$. Different colors correspond to different system sizes.}
    \label{fig:coll_spec}
\end{figure}
It is also important to understand if the DM is really generalizing from the data, thus producing new configurations, and it is not simply memorizing the dataset, and converging to one of the data points in the backward process. To answer in a quantitative way, we look at the distribution of the overlap between the generated data and the original data
\begin{equation}
    q^\mu = \frac{1}{N} \sum_{i=1}^N a_i^0 \text{ sgn}[x_i^\mu(t=0)] \; .
    \label{eq:overlap}
\end{equation}
As we have described before, because of the nature of the diffusion process, we generate real data and not binary data, and we eventually apply the sign function to make them binary again. In this way, if a generated configuration is present in the original training dataset, we will get a value of $q = 1$. In Fig.~\ref{fig:coll_spec} we show the histogram of the overlap $P(q)$. One can see that in the cold phase, it has two peaks that correspond to configurations that have aligned (positive peak) or opposite (negative peak) mean magnetization. There is no sizable peak at $q=1$, which implies that our model is not memorizing the training data, but it is always running a bona fide generalization. This is because the only parameters we are learning from the data are the correlation $\bm{C}_0$ and the magnetization, so the model is able to reproduce its critical properties which depend on the correlation of the data. As reported in \cite{biroli24}, to reduce the collapse of the backward process of the training data, an exponential number of training samples in $N$ is required. In modern applications, regularization can mitigate this problem \cite{generalization23}, and in our case, we can think that the use of this particular form of the score function acts as a regularization.

Our last, important goal is to characterize the time at which the trajectories of the backward process cluster and form two separate peaks in the $P(m)$. Following \cite{biroli24} we study the probability that two trajectories, cloned at time $t$, fall in the same cluster or peak at $t=0$, denoted by $\phi(t)$. In \cite{biroli24} the speciation time $t_s = \log(\Lambda)/2$ was introduced, where $\Lambda$ is the largest eigenvalue of $\bm{C}_0$. This proposal comes from the observation that close to $t_s$ the fluctuations of the data become of the order of the Gaussian noise.

In our approach, the relevant fluctuations leading to a clustering of the trajectories are expected to arise close to the time $t^*$ discussed above. We show in Fig.~\ref{fig:coll_spec} that the $\phi(t)$ curves do have a clean crossing when plotted as a function of $t/t^*$ and the slope at the crossing point increases with the system size $L$, indicating that, for $L\to\infty$, $t^*$ plays the role of a threshold. In \cite{biroli24} the same crossing has been found plotting $\phi(t)$ as a function of $t/t_s$, so both $t^*$ and $t_s$ belong to the critical region and define the same speciation time in the large $L$ limit. Indeed we find that both $t^*$ and $t_s$ grow logarithmically with $L$, while their difference stays almost constant.

\paragraph{Conclusions}
We have introduced a simple diffusion model that efficiently learns the high-dimensional probability distribution of configurations in a disordered model, namely the $2D$ diluted Ising model. Once trained, it can generate typical configurations above, below, and at the critical temperature, reconstructing accurately the magnetization and the critical exponents $\eta$ and $\nu$. Using this simple model we have better characterized the speciation time $t^*$ in terms of the effective Hessian $\bm{A}(t)$.

Our results clearly show that it is possible to use a diffusion model to reconstruct a statistical model with quenched disorder. This has potentially very relevant consequences that will be explored in the future. \\
\begin{acknowledgments}
We warmly thank Giulio Biroli and Marc M\'ezard for the interesting conversations and the critical reading of the manuscript. EM acknowledges the hospitality of Mark Bowick at UCSB (CA, USA), where part of this work was done, and the support by grant NSF PHY-2309135 to the Kavli Institute for Theoretical Physics (KITP). This research has been supported by funding from the 2021 FIS (Fondo Italiano per la Scienza) funding scheme (FIS783 - SMaC - Statistical Mechanics and Complexity), from the 2022 PRIN funding scheme (2022LMHTET - Complexity, disorder and fluctuations: spin glass physics and beyond) from Italian MUR (Ministry of University and Research) and from ICSC - Italian Research Center on High-Performance Computing, Big Data, and Quantum Computing, funded by the European Union - NextGenerationEU.
\end{acknowledgments}
\bibliographystyle{apsrev4-1} 
\bibliography{direbib}

\providecommand{\noopsort}[1]{}\providecommand{\singleletter}[1]{#1}%
\begin{thebibliography}{26}%
\makeatletter
\providecommand \@ifxundefined [1]{%
 \@ifx{#1\undefined}
}%
\providecommand \@ifnum [1]{%
 \ifnum #1\expandafter \@firstoftwo
 \else \expandafter \@secondoftwo
 \fi
}%
\providecommand \@ifx [1]{%
 \ifx #1\expandafter \@firstoftwo
 \else \expandafter \@secondoftwo
 \fi
}%
\providecommand \natexlab [1]{#1}%
\providecommand \enquote  [1]{``#1''}%
\providecommand \bibnamefont  [1]{#1}%
\providecommand \bibfnamefont [1]{#1}%
\providecommand \citenamefont [1]{#1}%
\providecommand \href@noop [0]{\@secondoftwo}%
\providecommand \href [0]{\begingroup \@sanitize@url \@href}%
\providecommand \@href[1]{\@@startlink{#1}\@@href}%
\providecommand \@@href[1]{\endgroup#1\@@endlink}%
\providecommand \@sanitize@url [0]{\catcode `\\12\catcode `\$12\catcode `\&12\catcode `\#12\catcode `\^12\catcode `\_12\catcode `\%12\relax}%
\providecommand \@@startlink[1]{}%
\providecommand \@@endlink[0]{}%
\providecommand \url  [0]{\begingroup\@sanitize@url \@url }%
\providecommand \@url [1]{\endgroup\@href {#1}{\urlprefix }}%
\providecommand \urlprefix  [0]{URL }%
\providecommand \Eprint [0]{\href }%
\providecommand \doibase [0]{http://dx.doi.org/}%
\providecommand \selectlanguage [0]{\@gobble}%
\providecommand \bibinfo  [0]{\@secondoftwo}%
\providecommand \bibfield  [0]{\@secondoftwo}%
\providecommand \translation [1]{[#1]}%
\providecommand \BibitemOpen [0]{}%
\providecommand \bibitemStop [0]{}%
\providecommand \bibitemNoStop [0]{.\EOS\space}%
\providecommand \EOS [0]{\spacefactor3000\relax}%
\providecommand \BibitemShut  [1]{\csname bibitem#1\endcsname}%
\let\auto@bib@innerbib\@empty
\bibitem [{\citenamefont {Sohl-Dickstein}\ \emph {et~al.}(2015)\citenamefont {Sohl-Dickstein}, \citenamefont {Weiss}, \citenamefont {Maheswaranathan},\ and\ \citenamefont {Ganguli}}]{sohl}%
  \BibitemOpen
  \bibfield  {author} {\bibinfo {author} {\bibfnamefont {J.}~\bibnamefont {Sohl-Dickstein}}, \bibinfo {author} {\bibfnamefont {E.}~\bibnamefont {Weiss}}, \bibinfo {author} {\bibfnamefont {N.}~\bibnamefont {Maheswaranathan}}, \ and\ \bibinfo {author} {\bibfnamefont {S.}~\bibnamefont {Ganguli}},\ }in\ \href {https://proceedings.mlr.press/v37/sohl-dickstein15.html} {\emph {\bibinfo {booktitle} {Proceedings of the 32nd International Conference on Machine Learning}}},\ \bibinfo {series} {Proceedings of Machine Learning Research}, Vol.~\bibinfo {volume} {37},\ \bibinfo {editor} {edited by\ \bibinfo {editor} {\bibfnamefont {F.}~\bibnamefont {Bach}}\ and\ \bibinfo {editor} {\bibfnamefont {D.}~\bibnamefont {Blei}}}\ (\bibinfo  {publisher} {PMLR},\ \bibinfo {address} {Lille, France},\ \bibinfo {year} {2015})\ pp.\ \bibinfo {pages} {2256--2265}\BibitemShut {NoStop}%
\bibitem [{\citenamefont {Yang}\ \emph {et~al.}(2023)\citenamefont {Yang}, \citenamefont {Zhang}, \citenamefont {Song}, \citenamefont {Hong}, \citenamefont {Xu}, \citenamefont {Zhao}, \citenamefont {Zhang}, \citenamefont {Cui},\ and\ \citenamefont {Yang}}]{comprehensiveSurvey}%
  \BibitemOpen
  \bibfield  {author} {\bibinfo {author} {\bibfnamefont {L.}~\bibnamefont {Yang}}, \bibinfo {author} {\bibfnamefont {Z.}~\bibnamefont {Zhang}}, \bibinfo {author} {\bibfnamefont {Y.}~\bibnamefont {Song}}, \bibinfo {author} {\bibfnamefont {S.}~\bibnamefont {Hong}}, \bibinfo {author} {\bibfnamefont {R.}~\bibnamefont {Xu}}, \bibinfo {author} {\bibfnamefont {Y.}~\bibnamefont {Zhao}}, \bibinfo {author} {\bibfnamefont {W.}~\bibnamefont {Zhang}}, \bibinfo {author} {\bibfnamefont {B.}~\bibnamefont {Cui}}, \ and\ \bibinfo {author} {\bibfnamefont {M.-H.}\ \bibnamefont {Yang}},\ }\href@noop {} {\bibfield  {journal} {\bibinfo  {journal} {ACM Computing Surveys}\ }\textbf {\bibinfo {volume} {56}},\ \bibinfo {pages} {1} (\bibinfo {year} {2023})}\BibitemShut {NoStop}%
\bibitem [{\citenamefont {Song}\ \emph {et~al.}(2020)\citenamefont {Song}, \citenamefont {Sohl-Dickstein}, \citenamefont {Kingma}, \citenamefont {Kumar}, \citenamefont {Ermon},\ and\ \citenamefont {Poole}}]{song2020score}%
  \BibitemOpen
  \bibfield  {author} {\bibinfo {author} {\bibfnamefont {Y.}~\bibnamefont {Song}}, \bibinfo {author} {\bibfnamefont {J.}~\bibnamefont {Sohl-Dickstein}}, \bibinfo {author} {\bibfnamefont {D.~P.}\ \bibnamefont {Kingma}}, \bibinfo {author} {\bibfnamefont {A.}~\bibnamefont {Kumar}}, \bibinfo {author} {\bibfnamefont {S.}~\bibnamefont {Ermon}}, \ and\ \bibinfo {author} {\bibfnamefont {B.}~\bibnamefont {Poole}},\ }\href@noop {} {\bibfield  {journal} {\bibinfo  {journal} {arXiv preprint arXiv:2011.13456}\ } (\bibinfo {year} {2020})}\BibitemShut {NoStop}%
\bibitem [{\citenamefont {Song}\ \emph {et~al.}(2021)\citenamefont {Song}, \citenamefont {Sohl-Dickstein}, \citenamefont {Kingma}, \citenamefont {Kumar}, \citenamefont {Ermon},\ and\ \citenamefont {Poole}}]{song21}%
  \BibitemOpen
  \bibfield  {author} {\bibinfo {author} {\bibfnamefont {Y.}~\bibnamefont {Song}}, \bibinfo {author} {\bibfnamefont {J.}~\bibnamefont {Sohl-Dickstein}}, \bibinfo {author} {\bibfnamefont {D.~P.}\ \bibnamefont {Kingma}}, \bibinfo {author} {\bibfnamefont {A.}~\bibnamefont {Kumar}}, \bibinfo {author} {\bibfnamefont {S.}~\bibnamefont {Ermon}}, \ and\ \bibinfo {author} {\bibfnamefont {B.}~\bibnamefont {Poole}},\ }\href {https://arxiv.org/abs/2011.13456} {\enquote {\bibinfo {title} {Score-based generative modeling through stochastic differential equations},}\ } (\bibinfo {year} {2021}),\ \Eprint {http://arxiv.org/abs/2011.13456} {arXiv:2011.13456} \BibitemShut {NoStop}%
\bibitem [{\citenamefont {Ho}\ \emph {et~al.}(2020)\citenamefont {Ho}, \citenamefont {Jain},\ and\ \citenamefont {Abbeel}}]{ddpmHo}%
  \BibitemOpen
  \bibfield  {author} {\bibinfo {author} {\bibfnamefont {J.}~\bibnamefont {Ho}}, \bibinfo {author} {\bibfnamefont {A.}~\bibnamefont {Jain}}, \ and\ \bibinfo {author} {\bibfnamefont {P.}~\bibnamefont {Abbeel}},\ }in\ \href {https://proceedings.neurips.cc/paper_files/paper/2020/file/4c5bcfec8584af0d967f1ab10179ca4b-Paper.pdf} {\emph {\bibinfo {booktitle} {Advances in Neural Information Processing Systems}}},\ Vol.~\bibinfo {volume} {33},\ \bibinfo {editor} {edited by\ \bibinfo {editor} {\bibfnamefont {H.}~\bibnamefont {Larochelle}}, \bibinfo {editor} {\bibfnamefont {M.}~\bibnamefont {Ranzato}}, \bibinfo {editor} {\bibfnamefont {R.}~\bibnamefont {Hadsell}}, \bibinfo {editor} {\bibfnamefont {M.}~\bibnamefont {Balcan}}, \ and\ \bibinfo {editor} {\bibfnamefont {H.}~\bibnamefont {Lin}}}\ (\bibinfo  {publisher} {Curran Associates, Inc.},\ \bibinfo {year} {2020})\ pp.\ \bibinfo {pages} {6840--6851}\BibitemShut {NoStop}%
\bibitem [{\citenamefont {Li}\ \emph {et~al.}(2024)\citenamefont {Li}, \citenamefont {Biferale}, \citenamefont {Bonaccorso}, \citenamefont {Scarpolini},\ and\ \citenamefont {Buzzicotti}}]{li24}%
  \BibitemOpen
  \bibfield  {author} {\bibinfo {author} {\bibfnamefont {T.}~\bibnamefont {Li}}, \bibinfo {author} {\bibfnamefont {L.}~\bibnamefont {Biferale}}, \bibinfo {author} {\bibfnamefont {F.}~\bibnamefont {Bonaccorso}}, \bibinfo {author} {\bibfnamefont {M.~A.}\ \bibnamefont {Scarpolini}}, \ and\ \bibinfo {author} {\bibfnamefont {M.}~\bibnamefont {Buzzicotti}},\ }\href@noop {} {\bibfield  {journal} {\bibinfo  {journal} {Nature Machine Intelligence}\ ,\ \bibinfo {pages} {1}} (\bibinfo {year} {2024})}\BibitemShut {NoStop}%
\bibitem [{\citenamefont {de~Haan}\ \emph {et~al.}(2021)\citenamefont {de~Haan}, \citenamefont {Rainone}, \citenamefont {Cheng},\ and\ \citenamefont {Bondesan}}]{dehaan21}%
  \BibitemOpen
  \bibfield  {author} {\bibinfo {author} {\bibfnamefont {P.}~\bibnamefont {de~Haan}}, \bibinfo {author} {\bibfnamefont {C.}~\bibnamefont {Rainone}}, \bibinfo {author} {\bibfnamefont {M.~C.}\ \bibnamefont {Cheng}}, \ and\ \bibinfo {author} {\bibfnamefont {R.}~\bibnamefont {Bondesan}},\ }\href@noop {} {\bibfield  {journal} {\bibinfo  {journal} {arXiv preprint arXiv:2110.02673}\ } (\bibinfo {year} {2021})}\BibitemShut {NoStop}%
\bibitem [{\citenamefont {Devlin}\ \emph {et~al.}(2024)\citenamefont {Devlin}, \citenamefont {Qiu}, \citenamefont {Ringer},\ and\ \citenamefont {Sato}}]{devlin24}%
  \BibitemOpen
  \bibfield  {author} {\bibinfo {author} {\bibfnamefont {P.}~\bibnamefont {Devlin}}, \bibinfo {author} {\bibfnamefont {J.-W.}\ \bibnamefont {Qiu}}, \bibinfo {author} {\bibfnamefont {F.}~\bibnamefont {Ringer}}, \ and\ \bibinfo {author} {\bibfnamefont {N.}~\bibnamefont {Sato}},\ }\href {\doibase 10.1103/PhysRevD.110.016030} {\bibfield  {journal} {\bibinfo  {journal} {Phys. Rev. D}\ }\textbf {\bibinfo {volume} {110}},\ \bibinfo {pages} {016030} (\bibinfo {year} {2024})}\BibitemShut {NoStop}%
\bibitem [{\citenamefont {Hunt-Smith}\ \emph {et~al.}(2023)\citenamefont {Hunt-Smith}, \citenamefont {Melnitchouk}, \citenamefont {Ringer}, \citenamefont {Sato}, \citenamefont {Thomas},\ and\ \citenamefont {White}}]{hunt23}%
  \BibitemOpen
  \bibfield  {author} {\bibinfo {author} {\bibfnamefont {N.~T.}\ \bibnamefont {Hunt-Smith}}, \bibinfo {author} {\bibfnamefont {W.}~\bibnamefont {Melnitchouk}}, \bibinfo {author} {\bibfnamefont {F.}~\bibnamefont {Ringer}}, \bibinfo {author} {\bibfnamefont {N.}~\bibnamefont {Sato}}, \bibinfo {author} {\bibfnamefont {A.~W.}\ \bibnamefont {Thomas}}, \ and\ \bibinfo {author} {\bibfnamefont {M.~J.}\ \bibnamefont {White}},\ }\href {https://arxiv.org/abs/2309.01454} {\enquote {\bibinfo {title} {Accelerating markov chain monte carlo sampling with diffusion models},}\ } (\bibinfo {year} {2023}),\ \Eprint {http://arxiv.org/abs/2309.01454} {arXiv:2309.01454 [hep-ph]} \BibitemShut {NoStop}%
\bibitem [{\citenamefont {Biroli}\ and\ \citenamefont {M\'ezard}(2023)}]{biroli23}%
  \BibitemOpen
  \bibfield  {author} {\bibinfo {author} {\bibfnamefont {G.}~\bibnamefont {Biroli}}\ and\ \bibinfo {author} {\bibfnamefont {M.}~\bibnamefont {M\'ezard}},\ }\href@noop {} {\bibfield  {journal} {\bibinfo  {journal} {J.\ Stat.\ Mech}\ ,\ \bibinfo {pages} {093402}} (\bibinfo {year} {2023})}\BibitemShut {NoStop}%
\bibitem [{\citenamefont {Ambrogioni}(2023)}]{ambrogioni2023}%
  \BibitemOpen
  \bibfield  {author} {\bibinfo {author} {\bibfnamefont {L.}~\bibnamefont {Ambrogioni}},\ }\href@noop {} {\bibfield  {journal} {\bibinfo  {journal} {arXiv preprint arXiv:2310.17467}\ } (\bibinfo {year} {2023})}\BibitemShut {NoStop}%
\bibitem [{\citenamefont {Raya}\ and\ \citenamefont {Ambrogioni}(2024)}]{raya2024spontaneous}%
  \BibitemOpen
  \bibfield  {author} {\bibinfo {author} {\bibfnamefont {G.}~\bibnamefont {Raya}}\ and\ \bibinfo {author} {\bibfnamefont {L.}~\bibnamefont {Ambrogioni}},\ }\href@noop {} {\bibfield  {journal} {\bibinfo  {journal} {Advances in Neural Information Processing Systems}\ }\textbf {\bibinfo {volume} {36}} (\bibinfo {year} {2024})}\BibitemShut {NoStop}%
\bibitem [{\citenamefont {Biroli}\ \emph {et~al.}(2024)\citenamefont {Biroli}, \citenamefont {Bonnaire}, \citenamefont {de~Bortoli},\ and\ \citenamefont {M\'ezard}}]{biroli24}%
  \BibitemOpen
  \bibfield  {author} {\bibinfo {author} {\bibfnamefont {G.}~\bibnamefont {Biroli}}, \bibinfo {author} {\bibfnamefont {T.}~\bibnamefont {Bonnaire}}, \bibinfo {author} {\bibfnamefont {V.}~\bibnamefont {de~Bortoli}}, \ and\ \bibinfo {author} {\bibfnamefont {M.}~\bibnamefont {M\'ezard}},\ }\href@noop {} {} (\bibinfo {year} {2024}),\ \Eprint {http://arxiv.org/abs/arXiv:2402.18491} {arXiv:2402.18491} \BibitemShut {NoStop}%
\bibitem [{\citenamefont {Ambrogioni}(2024)}]{AssociativeAL}%
  \BibitemOpen
  \bibfield  {author} {\bibinfo {author} {\bibfnamefont {L.}~\bibnamefont {Ambrogioni}},\ }\href {\doibase 10.3390/e26050381} {\bibfield  {journal} {\bibinfo  {journal} {Entropy}\ }\textbf {\bibinfo {volume} {26}} (\bibinfo {year} {2024}),\ 10.3390/e26050381}\BibitemShut {NoStop}%
\bibitem [{\citenamefont {Ghio}\ \emph {et~al.}(2023)\citenamefont {Ghio}, \citenamefont {Dandi}, \citenamefont {Krzakala},\ and\ \citenamefont {Zdeborov{\'a}}}]{ghio}%
  \BibitemOpen
  \bibfield  {author} {\bibinfo {author} {\bibfnamefont {D.}~\bibnamefont {Ghio}}, \bibinfo {author} {\bibfnamefont {Y.}~\bibnamefont {Dandi}}, \bibinfo {author} {\bibfnamefont {F.}~\bibnamefont {Krzakala}}, \ and\ \bibinfo {author} {\bibfnamefont {L.}~\bibnamefont {Zdeborov{\'a}}},\ }\href@noop {} {\bibfield  {journal} {\bibinfo  {journal} {arXiv preprint arXiv:2308.14085}\ } (\bibinfo {year} {2023})}\BibitemShut {NoStop}%
\bibitem [{\citenamefont {Anderson}(1982)}]{anderson82}%
  \BibitemOpen
  \bibfield  {author} {\bibinfo {author} {\bibfnamefont {B.~D.}\ \bibnamefont {Anderson}},\ }\href@noop {} {\bibfield  {journal} {\bibinfo  {journal} {Stochastic Processes and their Applications}\ }\textbf {\bibinfo {volume} {12}},\ \bibinfo {pages} {313} (\bibinfo {year} {1982})}\BibitemShut {NoStop}%
\bibitem [{\citenamefont {Griffiths}(1969)}]{griff69}%
  \BibitemOpen
  \bibfield  {author} {\bibinfo {author} {\bibfnamefont {R.~B.}\ \bibnamefont {Griffiths}},\ }\href@noop {} {\bibfield  {journal} {\bibinfo  {journal} {Physical Review Letters}\ }\textbf {\bibinfo {volume} {23}},\ \bibinfo {pages} {17} (\bibinfo {year} {1969})}\BibitemShut {NoStop}%
\bibitem [{\citenamefont {Harris}(1974)}]{harris74}%
  \BibitemOpen
  \bibfield  {author} {\bibinfo {author} {\bibfnamefont {A.~B.}\ \bibnamefont {Harris}},\ }\href@noop {} {\bibfield  {journal} {\bibinfo  {journal} {Journal of Physics C: Solid State Physics}\ }\textbf {\bibinfo {volume} {7}},\ \bibinfo {pages} {1671} (\bibinfo {year} {1974})}\BibitemShut {NoStop}%
\bibitem [{\citenamefont {Ballesteros}\ \emph {et~al.}(1997)\citenamefont {Ballesteros}, \citenamefont {Fernández}, \citenamefont {Martín-Mayor}, \citenamefont {Sudupe}, \citenamefont {Parisi},\ and\ \citenamefont {Ruiz-Lorenzo}}]{ballesteros97}%
  \BibitemOpen
  \bibfield  {author} {\bibinfo {author} {\bibfnamefont {H.~G.}\ \bibnamefont {Ballesteros}}, \bibinfo {author} {\bibfnamefont {L.~A.}\ \bibnamefont {Fernández}}, \bibinfo {author} {\bibfnamefont {V.}~\bibnamefont {Martín-Mayor}}, \bibinfo {author} {\bibfnamefont {A.~M.}\ \bibnamefont {Sudupe}}, \bibinfo {author} {\bibfnamefont {G.}~\bibnamefont {Parisi}}, \ and\ \bibinfo {author} {\bibfnamefont {J.~J.}\ \bibnamefont {Ruiz-Lorenzo}},\ }\href {\doibase 10.1088/0305-4470/30/24/006} {\bibfield  {journal} {\bibinfo  {journal} {Journal of Physics A: Mathematical and General}\ }\textbf {\bibinfo {volume} {30}},\ \bibinfo {pages} {8379} (\bibinfo {year} {1997})}\BibitemShut {NoStop}%
\bibitem [{\citenamefont {H.~Chau~Nguyen}\ and\ \citenamefont {Berg}(2017)}]{zecchina17}%
  \BibitemOpen
  \bibfield  {author} {\bibinfo {author} {\bibfnamefont {R.~Z.}\ \bibnamefont {H.~Chau~Nguyen}}\ and\ \bibinfo {author} {\bibfnamefont {J.}~\bibnamefont {Berg}},\ }\href {\doibase 10.1080/00018732.2017.1341604} {\bibfield  {journal} {\bibinfo  {journal} {Advances in Physics}\ }\textbf {\bibinfo {volume} {66}},\ \bibinfo {pages} {197} (\bibinfo {year} {2017})}\BibitemShut {NoStop}%
\bibitem [{\citenamefont {Wolff}(1989)}]{wolff89}%
  \BibitemOpen
  \bibfield  {author} {\bibinfo {author} {\bibfnamefont {U.}~\bibnamefont {Wolff}},\ }\href@noop {} {\bibfield  {journal} {\bibinfo  {journal} {Physical Review Letters}\ }\textbf {\bibinfo {volume} {62}},\ \bibinfo {pages} {361} (\bibinfo {year} {1989})}\BibitemShut {NoStop}%
\bibitem [{\citenamefont {Newman}\ and\ \citenamefont {Barkema}(1999)}]{newmanb99}%
  \BibitemOpen
  \bibfield  {author} {\bibinfo {author} {\bibfnamefont {M.~E.~J.}\ \bibnamefont {Newman}}\ and\ \bibinfo {author} {\bibfnamefont {G.~T.}\ \bibnamefont {Barkema}},\ }\href@noop {} {\emph {\bibinfo {title} {Monte Carlo methods in statistical physics}}}\ (\bibinfo  {publisher} {Clarendon Press},\ \bibinfo {address} {Oxford},\ \bibinfo {year} {1999})\BibitemShut {NoStop}%
\bibitem [{Note1()}]{Note1}%
  \BibitemOpen
  \bibinfo {note} {The Wolff algorithm has a small correlation time \cite {newmanb99} and to generate independent samples, we took measurements that were appropriately spaced in time.}\BibitemShut {Stop}%
\bibitem [{\citenamefont {Hadjiagapiou}(2011)}]{hadjia11}%
  \BibitemOpen
  \bibfield  {author} {\bibinfo {author} {\bibfnamefont {I.~A.}\ \bibnamefont {Hadjiagapiou}},\ }\href@noop {} {\bibfield  {journal} {\bibinfo  {journal} {Physica A: Statistical Mechanics and its Applications}\ }\textbf {\bibinfo {volume} {390}},\ \bibinfo {pages} {1279} (\bibinfo {year} {2011})}\BibitemShut {NoStop}%
\bibitem [{\citenamefont {Binder}(1981)}]{binder81}%
  \BibitemOpen
  \bibfield  {author} {\bibinfo {author} {\bibfnamefont {K.}~\bibnamefont {Binder}},\ }\href@noop {} {\bibfield  {journal} {\bibinfo  {journal} {Zeitschrift f{\"u}r Physik B Condensed Matter}\ }\textbf {\bibinfo {volume} {43}},\ \bibinfo {pages} {119} (\bibinfo {year} {1981})}\BibitemShut {NoStop}%
\bibitem [{\citenamefont {Kadkhodaie}\ \emph {et~al.}(2023)\citenamefont {Kadkhodaie}, \citenamefont {Guth}, \citenamefont {Simoncelli},\ and\ \citenamefont {Mallat}}]{generalization23}%
  \BibitemOpen
  \bibfield  {author} {\bibinfo {author} {\bibfnamefont {Z.}~\bibnamefont {Kadkhodaie}}, \bibinfo {author} {\bibfnamefont {F.}~\bibnamefont {Guth}}, \bibinfo {author} {\bibfnamefont {E.~P.}\ \bibnamefont {Simoncelli}}, \ and\ \bibinfo {author} {\bibfnamefont {S.}~\bibnamefont {Mallat}},\ }\href@noop {} {\bibfield  {journal} {\bibinfo  {journal} {arXiv preprint arXiv:2310.02557}\ } (\bibinfo {year} {2023})}\BibitemShut {NoStop}%
\end{thebibliography}%
\end{document}